\documentclass{article}

\usepackage[a4paper]{geometry}                    
\usepackage{xurl}                                 
\usepackage{dirtree}
\usepackage[backref=page]{hyperref}               
\hypersetup{
linktocpage,                                      
colorlinks  = true,                               
urlcolor    = cyan,
linkcolor   = red,
citecolor   = blue
}
\renewcommand*{\backref}[1]{}
\renewcommand*{\backrefalt}[4]{[%
\ifcase #1 Not cited.%
  \or Cited on page~#2.%
  \else Cited on pages #2.%
\fi]}

\usepackage{datetime}                             
  \newdateformat{monthonly}{\monthname[\THEMONTH]}
\usepackage{amssymb,amsthm,amsmath}               
\usepackage{graphicx}                             
\usepackage{floatrow}                             
\usepackage{booktabs}                             
\usepackage{paralist}                             
\usepackage{parskip}

\begin{document}

\author{Naisheng Liang and Alex Potanin\\
Australian National University~\footnote{\url{alex.potanin@anu.edu.au}}}
\title{Evaluating the Language-Based Security for Plugin Development}

\maketitle


\begin{abstract}
With the increasing popularity of plugin-based software systems, ensuring the security of plugins has become a critical concern. When users install plugins or browse websites with plugins from an untrusted source, how can we be sure that they do have any undesirable functions implicitly? In this research, we present a comprehensive study on language-based security mechanisms for plugin development. We aim to enhance the understanding of access control vulnerabilities in plugins and explore effective security measures by introducing a capability-based system. We also developed and evaluated test plugins to assess the security mechanisms in popular development environments such as IntelliJ IDEA and Visual Studio Code by utilising Java, JavaScript, and associated APIs and frameworks. We also explore the concept of capability-based module systems as an alternative approach to plugin security. A comparative analysis is conducted to evaluate the effectiveness of capability-based systems in addressing access control vulnerabilities identified in earlier sections. Finally, recommendations for improving plugin security practices and tools will be presented, emphasizing the importance of robust security measures in the ever-evolving landscape of software plugins.
\end{abstract}



\section{Introduction}

In this section, some fundamental concepts and knowledge about plugin development in the current software industry will be illustrated, and the reason why it is a popular topic in the modern software development area will also be demonstrated in subsection \ref{sec:1.1}. We will also discuss the security issues accompanying the emergence of plugins in subsection \ref{sec:1.2}. At last, we will indicate the outline of this thesis in subsection \ref{sec:1.3}

\subsection{Popular Fields: Plugin Development in Modern Software Ecosystems}
\label{sec:1.1}
In modern software ecosystems, plugin development has become an essential part of software engineering due to its flexibility, extensibility, and cost-effectiveness, particularly for popular platforms such as IntelliJ and VScode\cite{1}. Plugins are software components that extend the functionality of existing software by adding new features, capabilities, or services to enhance the user experience which allows developers to customize existing software systems or create new functionalities without altering the core codebase of the system\cite{2}. Plugins can be developed by third-party developers or even end-users, making them a critical element of the software development process. The concept of plugins has not been proposed recently; it has been around for several decades. However, with the advent of modern software ecosystems, the use of plugins has become more prevalent than ever\cite{3}. Modern software ecosystems are complex and dynamic, and they require a high degree of flexibility to adapt to changing user needs. Plugins provide this flexibility by allowing developers to add new features to an application without having to modify the core functionality of the application.

In recent years, plugin development has become an integral part of many popular software ecosystems, including web browsers, content management systems, and integrated development environments\cite{4}. For example, web browsers such as Google Chrome and Mozilla Firefox support a wide range of plugins that allow users to customize their browsing experience and add new functionality to the browser\cite{5}. Similarly, content management systems like WordPress and Drupal support a vast number of plugins that can be used to extend the functionality of the core system\cite{6}.

The popularity of plugin development can be attributed to several factors. First, plugins allow developers to build on top of existing software applications, which can significantly reduce development time and effort. Second, plugins can be used to add new functionality to an application without requiring changes to the core codebase, which can make it easier to maintain the application over time. Finally, plugins can be developed by third-party developers, which can help to create a vibrant ecosystem of plugins around a particular software application.

\subsection{Challenges in plugin development}
\label{sec:1.2}
Despite its many benefits, plugin development is not without its challenges. One of the most significant challenges facing plugin developers is security in particular this can lead to untrusted code. In plugin development, the concept of untrusted code refers to code that originates from external or unverified sources and may pose security risks to the system. Untrusted code can come in various forms, including user-submitted plugins, third-party libraries, or code obtained from the internet. While the integration of such code can enhance the functionality and versatility of plugins, it also introduces potential risks and challenges that developers must address\cite{7}.

One of the primary security challenges associated with untrusted code in plugin development is ensuring the integrity and authenticity of plugins. Since plugins are developed by third-party developers, there is a risk that they may contain malicious code or vulnerabilities that could be exploited by attackers. In addition, the development process of plugins may involve the use of third-party libraries or frameworks, which may also have security issues that could be exploited by attackers\cite{8}. Therefore, it is crucial to ensure that plugins are developed using secure coding practices and undergo rigorous testing to detect and fix any vulnerabilities.

Another significant security challenge in plugin development is the need to manage access control. Plugins can be granted varying levels of access to system resources, and any vulnerabilities in these access control mechanisms can be exploited by attackers to gain unauthorized access\cite{9}. Therefore, it is essential to implement proper access control measures to restrict plugin access to only the necessary resources and functions.

Moreover, plugins may have dependencies on other plugins or libraries, which can introduce security risks. These dependencies may contain vulnerabilities that could be exploited by attackers \cite{10}. As a result, it is essential to monitor and update plugins and their dependencies regularly to ensure they are secure and free of vulnerabilities.

\subsection{Thesis statement and outline}
\label{sec:1.3}

In spite of plugin development bringing numerous benefits, there are significant security challenges associated with it. Access control vulnerabilities are critical security issues in plugin development that can result in serious consequences, such as data breaches and system compromises. Therefore, the primary objective of this thesis is to evaluate the effectiveness of language-based security mechanisms in preventing access control vulnerabilities in plugin development by exploring the security challenges that arise from unexpected access to files, networks, and other parts of the application\cite{11}. We aim to investigate the current state of security in plugin development, the existing solutions, and their limitations, and also investigate capability-based module systems' potential to address these vulnerabilities.                            

\section{Background}\label{chap:background}

In this section, we will first demonstrate the background of developing plugins through different languages and elaborate on the relationships among them as well as the history of plugins development in subsection \ref{sec:2.1}. After that, we will introduce one of the most important aspects: IDE which is the basis of our experiments in subsection \ref{sec:2.2}. Additionally, we will discuss the access control vulnerabilities in plugin development including the principles and consequences of them in subsection \ref{sec:2.3}. Finally, we will introduce and briefly discuss popular security mechanisms from two levels in subsection \ref{sec:2.4}.

\subsection{Background of programming languages and plugin development}
\label{sec:2.1}

Plugins are an essential component in modern software ecosystems, allowing developers to extend the functionality of existing software. The development of plugins has been facilitated by the availability of a range of programming languages, each with its own strengths and weaknesses.

One of the most popular programming languages for developing plugins is Java, due to its platform independence, object-oriented design, and widespread adoption. Java plugins can be used across multiple platforms and operating systems, and can be easily distributed to end-users\cite{12}. The Gradle provides a standard framework for developing Java plugins, enabling developers to create reusable and interoperable components\cite{13}.

Other programming languages that are commonly used for plugin development include C++, Python, and JavaScript. C++ is popular for developing plugins that require high performance, such as those used in video games or graphics-intensive applications. Python is commonly used for scripting plugins, due to its ease of use and flexibility. JavaScript is widely used for web-based plugins, such as those used in web browsers\cite{14}.

The history of plugin development dates back to the early days of software development. One of the first instances of plugin development can be traced back to the Emacs editor in the 1970s\cite{15}. Emacs provided a framework for users to extend the editor's functionality through the use of Lisp scripts.

In the 1990s, plugins became more widespread with the advent of web browsers. Browsers such as Netscape Navigator and Internet Explorer allowed developers to create plugins that extended the browser's functionality, such as the Adobe Flash plugin for playing multimedia content\cite{16}.

Today, plugins are used in a wide range of applications, from productivity software to video games. The availability of a range of programming languages has made it easier for developers to create plugins for a variety of platforms and use cases.

\subsection{Integrated development environment(IDE)}
\label{sec:2.2}

An Integrated Development Environment (IDE) is a software application that provides comprehensive facilities for software development. IDEs generally consist of a source code editor, build automation tools, and a debugger\cite{17}. IDEs help developers write, compile, and debug their code more efficiently, with features like syntax highlighting, code completion, and error checking. IDEs also provide support for version control systems, testing frameworks, and other tools that help developers write better software.

There are several popular IDEs used in plugin development, including the IntelliJ platform and VS Code. The IntelliJ platform is a popular Java IDE developed by JetBrains. It provides a range of tools for Java developers, including support for multiple languages, debugging, and code completion.[17] The IntelliJ platform also includes a plugin development kit, which allows developers to create custom plugins for IntelliJ\cite{19}. It has a modular architecture that allows it to be extended using plugins, making it a popular choice for plugin development.

VS Code, on the other hand, is a lightweight, cross-platform code editor developed by Microsoft. It has quickly become one of the most popular code editors in the world, thanks to its ease of use and wide range of extensions. It supports a range of programming languages, including JavaScript, TypeScript, and Python\cite{20}. VSCode has a powerful extension system that enables developers to customize the IDE's features and functionality.[20] This makes it a popular choice for plugin development, especially for web and cloud-based applications.

Both IntelliJ and VS Code provide powerful tools for plugin development, but they also introduce new security challenges. For example, plugins can be a source of vulnerabilities in an IDE. If a plugin is not properly designed or implemented, it can provide a way for an attacker to gain access to sensitive information or perform unauthorized actions on the user's system. IDEs also introduce new attack surfaces, since they have access to a wide range of system resources and can interact with third-party libraries and services.

\subsection{Access control vulnerabilities in plugin development}
\label{sec:2.3}
\subsubsection{Principles of access control in software systems}

Access control is the process of managing the permission to access resources in a system. Access control mechanisms are typically based on two key concepts: authentication and authorization\cite{22}. Authentication is the process of verifying the identity of a user or system, while authorization is the process of granting or denying access to specific resources based on the user's role or permissions.

Access control vulnerabilities in plugins arise when developers fail to implement effective access control mechanisms, leading to unauthorized access and manipulation of data and functionality. Attackers can exploit access control vulnerabilities in plugins to gain elevated privileges, execute unauthorized actions, and access sensitive data. This kind of vulnerability can occur due to a variety of reasons, such as weak or default passwords, improper authentication mechanisms, and insecure communication channels\cite{23}. These vulnerabilities can be exploited by attackers to gain unauthorized access to sensitive data, modify or delete critical files, or execute malicious code on the system\cite{24}.

One common access control vulnerability in plugin development is inadequate input validation. Developers must ensure that all input received from external sources, such as user input, is thoroughly validated to prevent unauthorized access and manipulation. Failure to validate input can result in attacks such as injection attacks, where attackers inject malicious code into the plugin and execute it with elevated privileges.

Another access control vulnerability in plugin development is inadequate authentication and authorization mechanisms. Developers must ensure that plugins authenticate and authorize users properly before granting them access to sensitive data and functionality. Failure to implement proper authentication and authorization can result in attacks such as privilege escalation, where attackers gain access to higher levels of privilege than intended.

\subsubsection{ Consequences of access control vulnerabilities}

Access control vulnerabilities in plugins can have serious consequences on the security of software systems. Attackers can exploit these vulnerabilities to gain unauthorized access to sensitive data or system resources, modify or delete data, or execute arbitrary code.

One of the most common consequences of access control vulnerabilities in plugins is data theft. When an attacker gains unauthorized access to sensitive data, they can steal it for their own purposes or sell it on the black market. This can have serious implications for individuals, businesses, and even governments, as it can lead to identity theft, financial losses, and other types of fraud.

Another consequence of access control vulnerabilities in plugins is system compromise. Attackers can use these vulnerabilities to gain control of a system and execute arbitrary code, which can be used to launch further attacks or cause damage to the system. For example, an attacker could use a plugin vulnerability to install malware on a system, which could then be used to steal data, monitor user activity, or launch other types of attacks.

The consequences of access control vulnerabilities in plugins can be severe, leading to data theft, system disruption, and financial loss. Access control vulnerabilities in plugins have been exploited in several high-profile attacks and several high-profile security incidents have highlighted the importance of secure access control mechanisms in plugin development in recent years, such as the 2017 Equifax breach, where attackers exploited a vulnerability in a plugin to steal sensitive data of over 147.9 million customers\cite{25}.

\subsection{Two common levels of security mechanisms}
\label{sec:2.4}

Security is a major concern in plugin development, as plugins are often granted privileged access to a user's system and can potentially be exploited by attackers. Therefore, it is important to implement effective security mechanisms to prevent unauthorized access and ensure the integrity of the plugin.

There are various security mechanisms that can be implemented in plugin development. One approach is to use language-based security mechanisms, which are designed to prevent security vulnerabilities at the language level\cite{26}. This type of security mechanism involves the use of programming languages that have built-in security features\cite{27}. For example, languages such as Java have built-in security features that help prevent memory corruption and buffer overflow attacks and a built-in security manager that can be used to restrict access to resources or prevent malicious code execution\cite{28}. Similarly, C\# has a security model that allows developers to create secure applications by implementing code access security and role-based security. These languages also have automatic garbage collection, which helps prevent memory leaks and other memory-related security issues\cite{29,Potanin:2013:YIA:2486788.2486886,potanin:2004:ftfjp}.

Another approach is to use capability-based security, which is designed to prevent unauthorized access to resources by only allowing authorized parties to access them. Capability-based security relies on the principle of least privilege, which means that users and plugins should only be granted the minimum amount of access necessary to perform their tasks\cite{30}. This approach is often used in web browser extensions, where plugins are only given access to certain resources such as cookies or browser history\cite{31}. Capability-based security can be implemented using various techniques such as sandboxes, access control, and virtual machines. Sandboxing, for example, is a technique that involves isolating a plugin or a program from the rest of the system to prevent it from accessing sensitive resources or data\cite{32}. Access control, on the other hand, refers to the use of policies and rules to regulate access to resources and prevent unauthorized access\cite{33}. Virtual machines can also be used to implement capability-based security by providing a secure execution environment for the plugin or program\cite{34}.

In addition to language-based and capability-based security, there are also various other security mechanisms that can be implemented in plugin development. For example, plugins can be digitally signed to ensure their authenticity and integrity. Digital signatures are used to verify that a plugin has not been tampered with since it was signed and to ensure that it comes from a trusted source\cite{35}. Encryption and hashing can be used to secure sensitive data and communication channels

In this research, we will investigate how different languages and IDEs structure and establish their security mechanisms on the language level from multiple perspectives. Also, we will discuss how capability-based design can amend or avoid these kinds of issues caused by language-based design.


\section{Methodology}\label{chap:methodology}

\setcounter{secnumdepth}{3}
In this section, we will first provide an overview of our methodology to show the methods we used in the research in subsection \ref{sec:3.1}, and then we will also introduce the design of our future experiment in subsection \ref{sec:3.2}.

\subsection{Overview of methodology}
\label{sec:3.1}
The purpose of this research is to investigate the security risks associated with plugin development in integrated development environments (IDEs) and to explore potential solutions to mitigate these risks by employing a capability-based design. To achieve this, a research design was created to gather data on the access control vulnerabilities present in plugin development and to evaluate different security mechanisms that are used to address these vulnerabilities.

The literature review provided an overview of the current state of research in plugin development, access control vulnerabilities, and security mechanisms. This information was used to inform the experimental design and to guide the selection of appropriate data collection methods.

The data collection methods used in this research included investigative approaches. Investigate data was collected through actual deployments of our test plugins on personal computers and lab machines. 

The primary sources of data for this study were the IntelliJ plugins and the VScode extensions developed. The plugins and extensions were designed to highlight text, create new files randomly, read information from users’ file systems, and send pre-filled Google forms, and their code was thoroughly examined. We also recorded the entire progress from deployment to actual testing on the plugins and extensions. 

Secondary sources of data included existing literature on programming languages and plugin development, particularly related to security mechanisms. We also consulted technical documentation on the IntelliJ and VScode platforms, as well as other related sources of information.

Overall, this research provides valuable insights into the security risks associated with plugin development in IDEs and identifies potential solutions to mitigate these risks through capability-based design.

\subsection{Future experiment design}
\label{sec:3.2}

Since this is a one-semester project, we have applied for the ARIES account and we are creating a new ethical protocol.  After that, we will apply for ethical approval from ANU, hence we obtained all data and results from our own devices. We also propose a more comprehensive experiment after we can be granted ethical approval from ANU in the future.

The data collection methods that will be used in future research include both investigative and quantitative approaches. Investigative data will be collected through actual deployments and use experiences of our test plugins by developers and other ordinary users without any background in computer science, while quantitative data will be gathered through surveys and questionnaires. The collected data will be then analyzed using statistical methods and quantitative analysis techniques to identify common access control vulnerabilities in plugin development and to evaluate the effectiveness of different security mechanisms.

To ensure the validity and reliability of the experimental design, randomized controlled trials will be conducted with different experimental groups. Participants will be randomly assigned to experimental groups and blinded to the treatment conditions to minimize the risk of bias. The ethical considerations associated with the research will also be taken into account, including informed consent procedures, potential harm to participants, and confidentiality procedures.

The limitations of the research included the small sample size of participants and the limited scope of the study. These limitations can be addressed by ensuring that the experimental design and data collection methods were rigorously applied, and by highlighting the need for further research in this area. 

\subsubsection{Data Collection}

In this section, we will describe the data collection methods used in this study. The data collected is crucial to the analysis of the effectiveness of the security mechanisms employed by the plugins developed. The sources of data are discussed below.

\paragraph{Sources of data}
The sources of data are the same as in the previous section we mentioned before. It has two sources mainly which are developed plugins and existing literature on programming languages.

\paragraph{Data collection methods} 
To collect data on the plugins and extensions, we will employ various methods, including code analysis, surveys, interviews, and user comments.

Code analysis involved examining the source code of the plugins and extensions to determine how they functioned and to identify any security vulnerabilities. This process will be carried out using static analysis methods to detect any issues in the code.

Surveys will be used to collect feedback from users of the plugins and extensions. We design the surveys to gather information on the ease of use, usefulness, and security of the plugins and extensions.

User comments will be collected from online forums and other sources of information on the plugins and extensions. The comments will be analyzed to identify any issues that users have encountered and to gauge their overall satisfaction with the plugins and extensions.

\paragraph{Data analysis}
The data collected will be analyzed using a combination of qualitative and quantitative methods. The qualitative analysis involves categorizing the data based on the themes and patterns that emerged from the surveys, questionnaires, and user comments. Quantitative analysis involves analyzing the data collected through surveys to identify any significant trends or patterns.

The analysis of the data collected allows us to identify any potential security vulnerabilities in the plugins and extensions and to assess the effectiveness of the security mechanisms employed. This information can be used to improve the security of the plugins and extensions and to develop new security mechanisms where necessary.

\subsubsection{Experiment Design}

The experiment design is a crucial aspect of this research, as it aims to evaluate how the security mechanisms work in plugin development with different languages. In this section, we provide an overview of the experiment design, including the experimental groups, randomization, and blinding procedures.

\paragraph{Overview of the Experiment Design}
The experiment design involves the use of three different groups: the control group, the experimental group 1, and the experimental group 2. The control group will be assigned the basic version of the plugin which only contains the function of highlighting without any harmful functions, whereas the experimental group 1 will also use the highlighter plugin developed in this research with additional functions of manipulating files like creating or reading files from their system. Experimental group 2 will use another plugin developed in this research, which not only can highlight words with customized colours as others but also access the network implicitly such as by sending a pre-filled Google form.

Participants will be randomly assigned to each group, and the experiment will be conducted on the IntelliJ and VScode platforms. The participants will be given a set of tasks, and their progress will be monitored to evaluate any security mechanisms that will be triggered and the awareness of access control vulnerabilities.

\paragraph{Description of the Participants and Volunteers}
The entire experimental and control group consists of 18 participants who will be provided with the developed plugins. These participants have different backgrounds and knowledge of computer science and security. Some of them never or seldom touch software development, some of them have a weak background in software engineering such as university students, and the rest of them are experienced developers or engineers who are experts in security and software development.

\paragraph{Explanation of Randomization and Blinding Procedures}
To ensure the validity of the results, the participants will be randomly assigned to each group. Randomization helps to eliminate the possibility of selection bias and ensures that each group is similar in terms of their abilities and expertise. The blinding procedure will be used to ensure that the participants are not aware of which group they have been assigned to, to avoid any potential biases.

\subsubsection{Perspectives from Other Researchers}

Several researchers have conducted similar experiments to evaluate the effectiveness of security mechanisms in plugin development. For example, \cite{36} evaluated the effectiveness of a security plugin that checks for security vulnerabilities in the code while developing WordPress plugins. Their results showed that the security plugin was effective in detecting security vulnerabilities in the code.

In another study, \cite{37} evaluated the effectiveness of security mechanisms in preventing security breaches in IoT systems. They used a randomized controlled trial to evaluate the effectiveness of their security mechanisms and found that their approach was effective in enhancing the security of IoT systems.

\subsubsection{Ethical Considerations}

Ethical considerations are an important aspect of any research study. In this section, we discuss the ethical considerations of our study, including informed consent procedures, potential harm to participants, and confidentiality procedures.

\paragraph{Informed Consent Procedures}
Informed consent is an essential component of ethical research. Participants must be fully informed of the study's purpose, procedures, and any potential risks or benefits before they can agree to participate. In our study, we will follow standard procedures for obtaining informed consent. We will provide potential participants with a consent form that details the purpose and procedures of the study, as well as any potential risks or benefits. We will also explain that participation was voluntary and that they could withdraw from the study at any time without penalty.

\paragraph{Potential Harm to Participants}
It is crucial to consider the potential harm that participants may experience when conducting a research study. We will conduct a thorough risk assessment and determine that there were actually no risks to our participants. The plugins we provide will not leak any personal information from participants. The study described involves only a survey including questionnaires and did not involve any physical or emotional stressors. However, we still took steps to ensure that participants were not subjected to any undue stress or discomfort during the study. We also make sure to offer support and resources to participants if they experience any emotional distress as a result of the study.

\paragraph{Confidentiality Procedures}
Confidentiality is another essential aspect of ethical research. We will take measures to ensure that participants' data remain confidential and that their privacy is protected. We will assign unique identification numbers to participants, and all data will be stored anonymously. All data will also be destroyed after the research is completed. We also make sure that only authorized personnel have access to the data.

\paragraph{Other Perspectives}
Several researchers have emphasized the importance of ethical considerations in research involving human subjects. According to a study by \cite{38} and her colleagues, researchers must follow ethical guidelines to ensure that participants are protected from harm and their rights are respected. Another study by \cite{39} highlights the need for informed consent and confidentiality in research involving human subjects.

Similarly, \cite{40} argues that researchers should prioritize respect for persons, beneficence, and justice when designing research studies. They also emphasize the importance of transparency and open communication with participants.

\subsubsection{Limitations}

As with any research, there are limitations to this study that should be taken into consideration.
This subsection will discuss the limitations of the current study and how they were addressed.

The first limitation is related to the sample size, as the study only included a small number of participants. A larger sample size would have been more representative and may have led to more accurate results. Additionally, the participants were recruited from a single location, which limits the generalizability of the findings to other populations.

Another limitation is related to the self-reported data collected in this study. Self-reported data can be subject to bias and may not accurately reflect participants’ actual behaviours or experiences. To mitigate this limitation, efforts were made to ensure that participants understood the questions being asked and that they felt comfortable providing honest answers. However, it is still possible that some participants may have provided inaccurate or incomplete responses.

A third limitation was the potential for confounding variables. Confounding variables are variables that can influence the outcome of the study but are not measured or controlled for. To address this limitation, the study used a randomized controlled design, which helps to control for confounding variables.
Finally, it is important to acknowledge the limitations related to the experimental design of this study. While efforts were made to minimize bias through randomization and blinding procedures, it is still possible that other factors may have influenced the results. 

Despite these limitations, this study provides valuable insights into how security mechanisms work in different plugin development environments with various languages. Future studies should aim to build on these findings by addressing the limitations identified and further exploring the factors that contribute to potential solutions to improve security mechanisms.                             

\section{Design and Implementation}
\label{chap:design and implementation}

In this section, we will elaborate on the aims and functions of each plugin developed briefly in subsection \ref{sec:4.1} firstly. After that, the specific workflow and designs will be stated in subsection \ref{sec:4.2}. The next subsection \ref{sec:4.3} will demonstrate the structures of plugins in different IDEs, how they are implemented, and involved frameworks as well as APIs. In subsection \ref{sec:4.4}, we show the physical properties of our test device. All experimental results will be presented in subsection \ref{sec:4.5}. In the last of this chapter, some limitations will be discussed briefly in subsection \ref{sec:4.6} 

\subsection{A brief description of plugins}
\label{sec:4.1}
In this section, we will provide a description of the plugin architecture used for the highlighter plugin, and how it was adapted for the two additional plugins with malicious functionality, namely the highlighter with file manipulator and the highlighter with network manipulator. The plugin architecture used in this project was designed to be modular and extensible, allowing for the easy addition of new features and functionality.

The highlighter plugin provides a simple interface for highlighting text in the application. It consists of a single class that implements the necessary interfaces for the plugin architecture. The highlighter plugin is activated when the user selects text in the interface of the IDE and right clicks the word and clicks highlight in the pop-up menu. Once activated, the plugin adds a highlight to the selected text as well as all other exact same words in the file and notifies the plugin manager of the change.

The highlighter with file manipulator plugin extends the functionality of the highlighter plugin by allowing the user to save highlighted text to a file. This plugin consists of two classes: one for handling the file operations and one for handling the highlighter functionality. The file manipulator class is responsible for opening, reading, and writing files, while the highlighter class is responsible for handling the highlighting of text. When the users execute the highlighting function, the file manipulator class is also called to open the file, write the arbitrary texts, save the file, or even create a brand new file.

The highlighter with network manipulator plugin extends the functionality of the highlighter plugin by allowing the user to send a highlighted text to a remote server. This plugin consists of two classes: one for handling the network operations and one for handling the highlighter functionality. The network manipulator class is responsible for opening a connection to the remote server, sending the information we set before as Google Forms, and receiving a response. The highlighter class is responsible for handling the highlighting of text. When the user clicks the send button, the network manipulator class is called to send any data that may be retrieved from users’ computers or is pre-filled to the remote server.

\begin{figure*}
   \label{fig:basic-highlighter}]{\includegraphics[width=10cm]{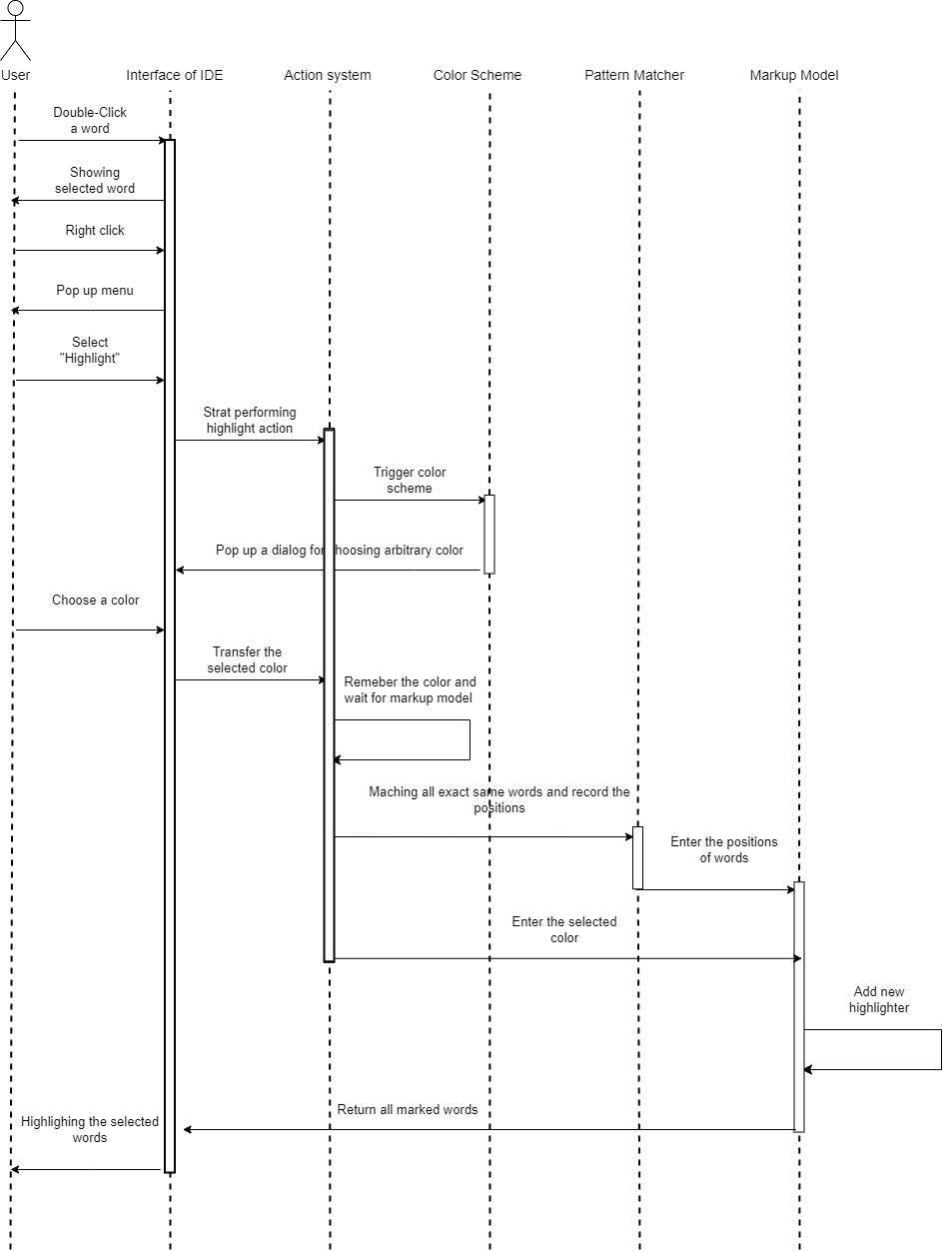}}
  \caption{Workflow of basic highlighter}
\end{figure*}

\subsection{Design and workflow of plugins}
\label{sec:4.2}

The implementation of the three plugins involved the use of two different programming languages and development environments mainly which are IntelliJ with Java and VScode with Javascript, as well as the integration of various libraries and frameworks. However, the essential design and workflow for each of the plugins are the same exactly, in this subsection, we will introduce their design and how they work.

For the basic highlighter plugin, shown in Figure~\ref{fig:basic-highlighter}, we aimed to develop it as other common highlighters in the marketplace which is supposed to have one simple function. The interactions with users should keep as easy and handy as possible. In our design, the user can choose any words or sentences by double-clicking or selecting the sentence, then right-clicking selected words, there will be a pop-up menu in which an additional option will be shown named “Highlight”. When users click this option, the highlighting procedure will be triggered, and a dialogue that contains three different colour schemes will pop up for users to choose any colour they want. After the colour is confirmed, the markup model will mark all words which have the exact same pattern as selected words as shown in Figure \ref{fig:basic-highlighter},

\begin{figure*}
   \label{fig:file-manipulator-1}]{\includegraphics[width=8cm]{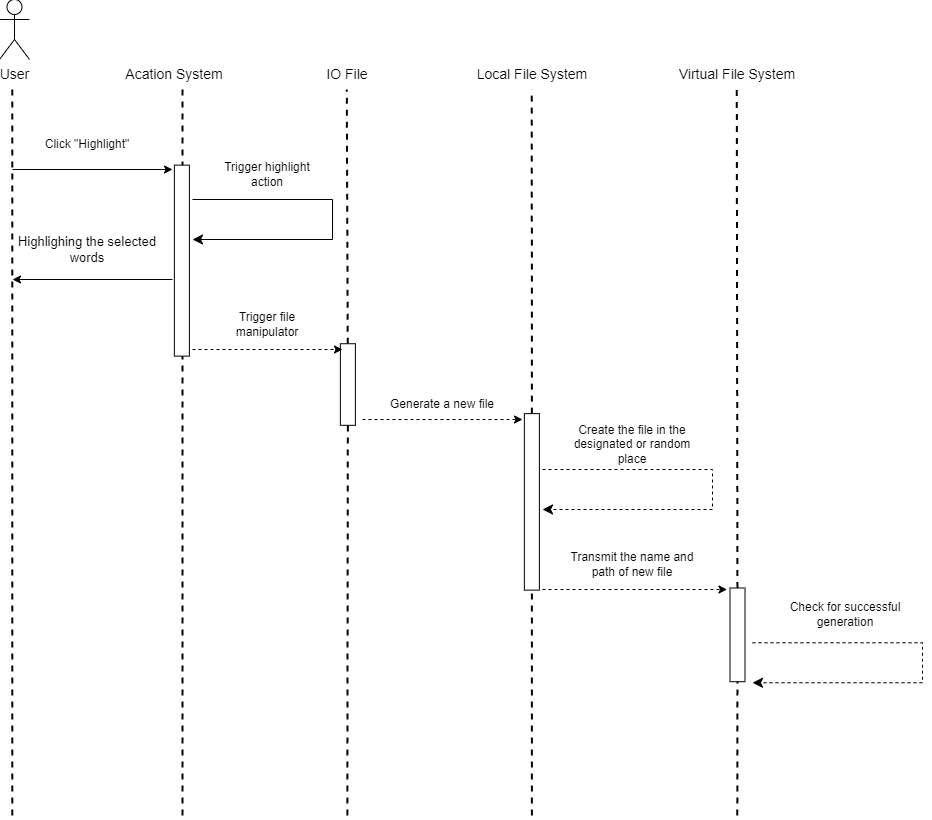}}
  \caption{Workflow of basic highlighter with file manipulator }
\end{figure*}

For the highlighter with file manipulator, the workflow is illustrated in Figure \ref{fig:file-manipulator-1}, it has another malicious function that can access to user’s file system and execute operations that we defined implicitly. To show them intuitively, we used dotted lines to present the implicit operations and simplify the basic highlighter function. To be more specific, when users click the highlight option, the hidden file manipulator will also be triggered. It can first generate a new file containing arbitrary data by IO File and employ the local file system to find the path where this file should be created. After the operation of creation is finished, it also invokes the virtual file system (VFS) to examine whether this file is created successfully or not.

\begin{figure*}
   \label{fig:network-manipulator}]{\includegraphics[width=8cm]{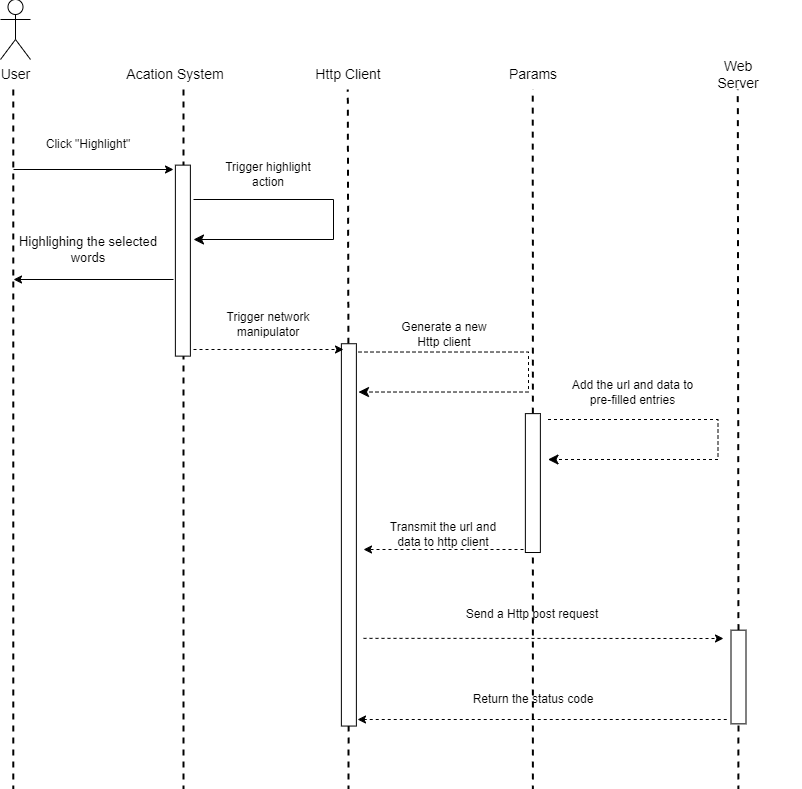}}
  \caption{Workflow of basic highlighter with network manipulator  }
\end{figure*}

The operating mechanism and workflow are quite similar to the previous plugin for the highlighter with network manager. As shown in Figure \ref{fig:network-manipulator}, the plugin will also trigger another implicit function that can send a post request to the web server. More specifically, it will generate an HTTP client automatically and then set the necessary parameters for all entries that we defined. After that, the HTTP client will send a post request with data that can be pre-filled or retrieved from the user’s computers to Google Forms so that we can collect them by checking the responses in the Google form.

\subsection{Implementation of the plugins}
\label{sec:4.3}

The implementation process of the plugins involved the use of programming languages and development environments to bring the desired functionalities to life. Each plugin was implemented with specific considerations and challenges. In this section, we will delve into the implementation process of the plugins, discussing the programming languages, development environments, and the challenges faced during the implementation.

\subsection{IntelliJ Development}

The first three plugins were implemented in IntelliJ IDEA, an integrated development environment (IDE), were implemented using Java and the IntelliJ Platform API. Java is a widely used programming language known for its versatility and compatibility. The IntelliJ Platform API provided the necessary tools and libraries for developing plugins specifically tailored for IntelliJ IDEA.

To be more specific, here the tree structure below shows a sample of the general structure of an IntelliJ plugin and Gradle build system:

\dirtree{%
.1 .
.2 build.gradle \DTcomment{Build script file}.
.2 gradle \DTcomment{Gradle wrapper}.
.3 wrapper.
.4 gradle-wrapper.jar.
.4 gradle-wrapper.properties \DTcomment{Wrapper configuration file}.
.2 gradlew \DTcomment{Shell script}.
.2 gradlew.bat \DTcomment{Batch script}.
.2 settings.gradle \DTcomment{Optional settings file for a single project}.
.2 src.
.3 main.
.4 java.
.5 io.
.6 org.
.7 example.
.8 plugin.
.9 Highlighter.java.
.9 FileManipulator.java.
.9 NetworkManipulator.java.
.4 resources.
.5 images.
.6 a.svg.
.6 b.svg.
.5 META-INF.
.6 plugin.xml \DTcomment{Plugin configuration file}.
.6 pluginIcon.svg \DTcomment{Plugin Logo, displayed in IDEA/Plugins}.
}

IntelliJ Plugin Structure:
An IntelliJ plugin follows a specific directory structure to organize its components. Here is a general overview:

1.	Plugin Root:
Contains the main plugin configuration file (plugin.xml) and other resources.
Additional directories and files for the plugin's functionality can be created here.

2.	Source Directory:
Contains the source code of the plugin, typically organized in packages.
Java files (.java) or Kotlin files (.kt) reside here, implementing the plugin's logic.

3.	Resources Directory:
Holds non-source code files, such as images, property files, and other resources.
It can include directories for specific resource types, like "images" or "messages" for localization.

Gradle Build System:

Gradle is a popular build automation tool used for building and managing projects. It simplifies the build process and handles dependencies efficiently\cite{41}. Here is an overview of a typical Gradle project structure:

1. build.gradle:
The build script file defines project-specific configurations, dependencies, and tasks.
It includes settings like plugin dependencies, build settings, and other project-specific configurations.

2. gradle Directory:
Contains the Gradle wrapper, which provides a self-contained Gradle distribution for the project.
It includes the "wrapper" directory with the Gradle wrapper JAR file and its properties file.

3. gradlew and gradlew.bat:
Shell and batch scripts respectively, are used to execute Gradle commands in a platform-independent manner.
They allow users to run Gradle tasks without requiring a pre-installed Gradle distribution.

4. settings.gradle:
An optional file is used to define project-specific settings, such as the project name, modules, and subprojects.
It is commonly used for multi-project setups to specify the project structure.
These structures provide a foundation for developing IntelliJ plugins using Gradle as the build system. They ensure organization and proper integration within the IntelliJ IDEA platform.

The first one is a highlighter plugin, which is the foundation for the other two plugins and was implemented using the Java programming language and the IntelliJ IDEA development environment. The plugin leveraged the IntelliJ platform's APIs for interacting with the editor, selecting the text, and applying markup highlighting. This choice of programming language and development environment was ideal for creating a plugin that seamlessly integrated with the IntelliJ IDEA IDE. It allowed for direct access to the editor and other necessary components of the IDE, enabling efficient highlighting of selected text. The implementation process involved understanding the API documentation, studying code samples, and experimenting with different approaches to achieve the desired functionality. 

When adapting the highlighter plugin for the file manipulator functionality, additional implementation steps were required. This involved integrating file system operations, such as retrieving specified files and creating a new file, into the existing plugin. The Java File and I/O APIs were utilized to manipulate files while maintaining compatibility with the IntelliJ platform. Careful error handling and permissions management were implemented to ensure the security and proper functioning of the plugin. Here we encountered a problem that some operating systems such as macOS have a special mechanism named system integrity protection (SIP) which is used to protect critical system files and directories from unauthorized modifications, even by users with administrative privileges. It is a security feature introduced in OS X El Capitan (version 10.11) and later versions of macOS\cite{42}. It restricts both system processes and third-party applications from modifying specific protected areas of the file system, including certain system directories, system files, and kernel extensions. In our experiments, we cannot create or modify any files in the root folder of the macOS system.

Similarly, for the network manipulator functionality, the highlighter plugin was extended to interact with a Google Form for data submission. This plugin included the capability to send data to a Google Form with pre-filled information. The implementation involved using HTTP client libraries and APIs to make requests to the Google Form endpoint and send the form data. This required integrating HTTP client libraries, such as Apache HttpClient, to send POST requests to the Google Form's URL with the pre-filled information. Proper handling of network connectivity, error responses, and anonymity considerations was addressed during the implementation.

Throughout the implementation process, various challenges were encountered and resolved. One challenge was ensuring compatibility and proper integration with the IntelliJ platform's APIs. This required a deep understanding of the platform's architecture and documentation. Additionally, handling exceptions and errors gracefully to provide a consistent experience for users without being aware of anything special was a significant consideration during implementation.

\subsection{VS Code Development}

For the Visual Studio Code (VS Code) extension, the plugin was implemented using JavaScript, which is the primary language for VS Code extensions. JavaScript offers extensive capabilities for developing extensions and interacting with the VS Code API. Additionally, Node.js, a JavaScript runtime environment, was used to execute the extension code. Here is a tree structure that represents a typical setup for a VS Code extension, with the necessary files and directories to define the plugin's functionality, configuration, and documentation.

\dirtree{%
.1 /.
.2 .vscode.
.3 launch.json \DTcomment{Configuration for plugin loading and debugging}.
.3 extensions.json \DTcomment{Configuration for JavaScript compilation tasks}.
.2 .gitignore \DTcomment{Ignore build output and node\_modules files}.
.2 README.md \DTcomment{A friendly plugin documentation}.
.2 src.
.3 extension.ts \DTcomment{Plugin source code}.
.2 package.json \DTcomment{Plugin configuration manifest}.
.2 jgconfig.json \DTcomment{JavaScript configuration}.
}

1 .vscode:
Contains configuration files specific to the VS Code extension.
launch.json specifies the configuration for plugin loading and debugging.
extensions.json provides configuration for JavaScript compilation tasks.

2 .gitignore:
Specifies files and directories to be ignored when using version control (such as Git).
Typically includes build output files and the node\_modules directory.

3. README.md:
A markdown file that serves as the documentation for the plugin.
It provides instructions, explanations, and other helpful information about the plugin.

4. src:
Contains the source code for the extension.
In this case, the extension.ts file represents the main source code file for the plugin.

5. package.json:
A JSON file that serves as the configuration manifest for the extension.
It includes metadata about the extension, such as name, version, dependencies, and entry points.

6. jgconfig.json:
A JSON file that provides JavaScript configuration settings.
It may include specific settings for the JavaScript code used in the extension.

In the actual implementation, we chose another simple function which is a file state checker as the appearance of our extension. Users can right-click the exact file and choose the check file option, then a dialogue will pop up which shows the information of this file such as the data of creation, last modification, and the size of this file. It is also integrated with two similar malicious functions as IntelliJ plugins. The file manipulator is implemented by the original APIs from VS code which are path and fs which can interact with the VS Code editor and file system, allowing the plugin to read file metadata, create new files, and perform other file-related operations. The network manipulator was implemented as a Visual Studio Code extension using JavaScript, along with additional libraries such as 'isomorphic-fetch' for making HTTP requests. This implementation allowed the plugin to send data to external endpoints, such as submitting form data to a Google Form. The use of JavaScript and the provided libraries facilitated communication with remote servers and the handling of HTTP requests.

\subsection{Experimental Setup}
\label{sec:4.4}

Table ~\ref{tab:1} shows the hardware platform of our test machine.
\begin{table*}[h]
{
\sffamily

\begin{tabular}{r@{\hspace{1.5ex}}r@{\hspace{1.5ex}}r@{\hspace{1.5ex}}}
\\[-2ex]
{\textbf{Memory}} & \multicolumn{1}{r@{\hspace{1.5ex}}}{\textbf{CPU}} &  \multicolumn{1}{r@{\hspace{1.5ex}}}{\textbf{Operating System}} \\
\midrule 
{\textbf{16 GB DDR4 2300 MHz }} & Intel Core i3 6-Cores 1.5 GHz & MacOS\\

\bottomrule
\end{tabular}
}
  \caption{The configuration of our test machine.}
  \label{tab:1}
\end{table*}

Software Configuration:
The development and execution of the plugins were performed using the following software components:

Integrated Development Environment (IDE): IntelliJ IDEA Community Edition version 2021.3.2

Programming Languages: Java for the IntelliJ IDEA plugins and JavaScript for the VSCode extension

Libraries and Frameworks: The IntelliJ IDEA plugins utilized the JetBrains Platform SDK 11 and Gradle development kit, while the VSCode extension relied on the VS code module for VSCode extension development.

Environmental Variables:
To ensure reliable and consistent results, several environmental variables were controlled during the experiment. These variables include:

Operating System Configuration: The experiments were conducted on the same version of macOS to eliminate any variations caused by different OS versions or configurations.

Development Environment Setup: The plugins were developed and tested in identical development environments, ensuring consistency in the codebase and development processes.

Plugin Configuration: The plugins were configured with default settings and options, ensuring a consistent experience across different testing scenarios.
By controlling these environmental variables, the study aimed to minimize any external factors that could introduce biases or affect the performance of the plugins.

\subsection{Testing and experiment results}
\label{sec:4.5}

After the complete implementation and deployment of our plugins, we test them on personal computers in three different ways.

The primary objective of the test is if any alerts or warnings will be triggered when users want to install or even use these plugins. If there are alerts triggered, what information do they present, and what actions they will perform? To achieve this, we prepared three ways to install the plugin and observe how they perform their tasks respectively which cover all methods of installing.

The first way is that users will be assigned a snapshot of the plugin which can be installed through the local computer system to the IDE, the plugin will be in effect at once. The second way is that users can also search for the name of the plugin on the marketplace and install it since we have already published them as other common plugins. The last way is that users can download the entire project to the local computer and run the plugin in the sandbox which is an isolated environment for testing provided by IDE itself.

As a result of our observation, when users try to use the Highlighter plugins containing other manipulators, all malicious functions work as we expected. None of them raise alerts or warnings for malicious functions and the users’ file system is modified as well as users’ network is also invaded. They exhibited accurate functionality, successfully highlighting the selected words, manipulating files, and interacting with external resources without any blocks. Furthermore, the testing also revealed that the plugins can gain almost the same rights to access any part of the computer as the administrator namely users. The plugins can retrieve and modify any files except the root folder protected by SIP which is the same as users.

\begin{figure*}
   \label{fig:file-manipulator-ij}]{\includegraphics[width=8cm]{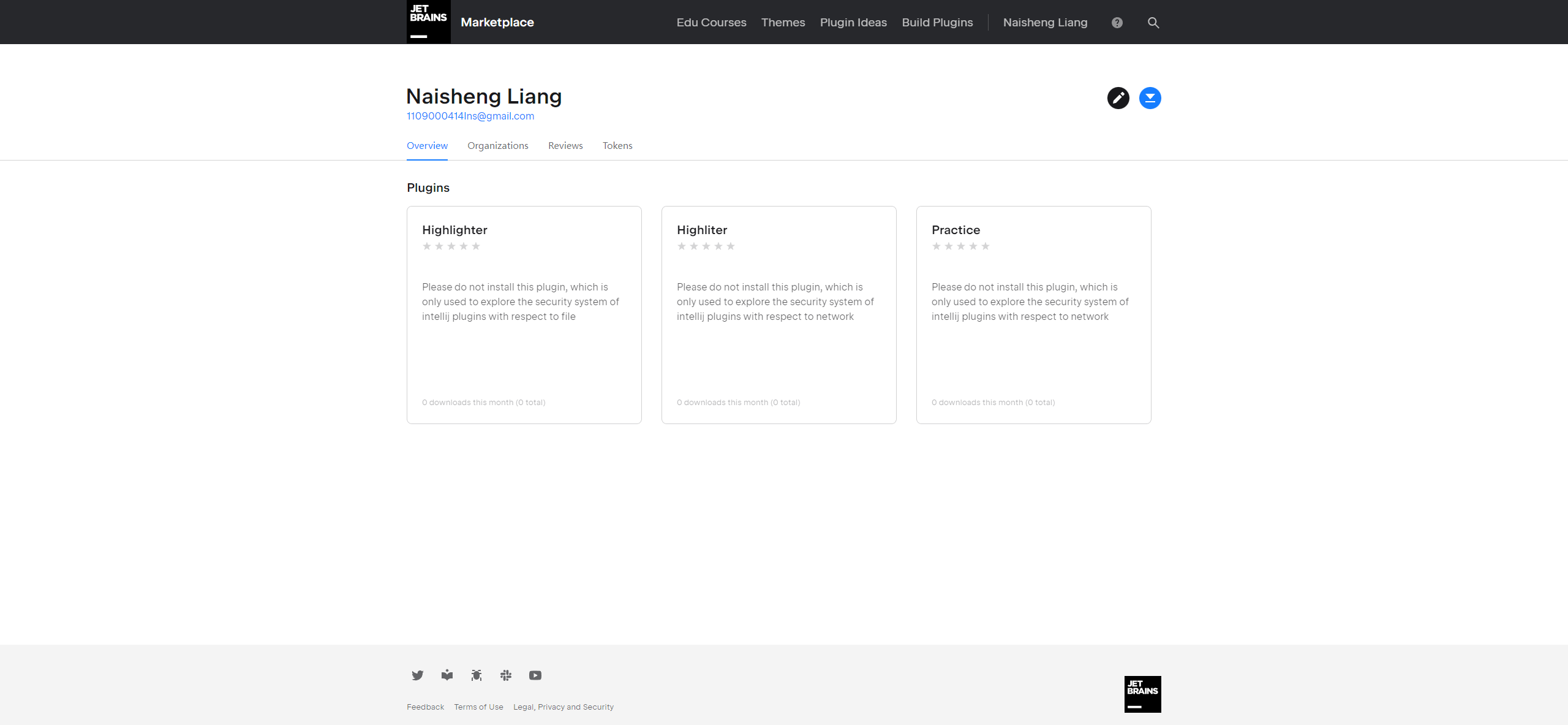}}
  \caption{Publishing of IntelliJ plugins}
\end{figure*}

\begin{figure*}
   \label{fig:file-manipulator-vs}]{\includegraphics[width=8cm]{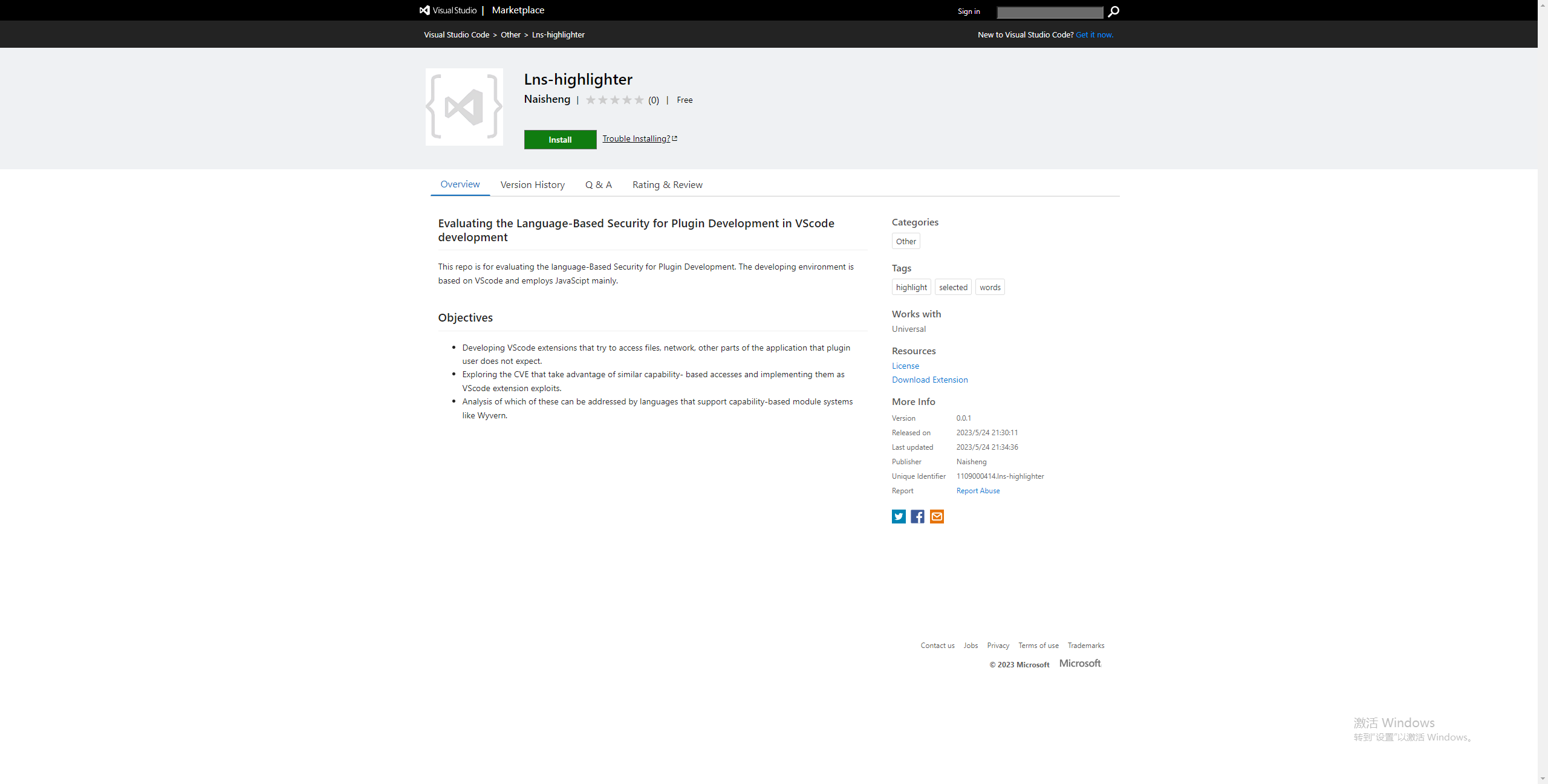}}
  \caption{Publishing of VS Code extensions}
\end{figure*}

\subsection{Limitations}
\label{sec:4.6}

While the design and implementation of the plugins and the experimental setup were carefully planned and executed, it is important to acknowledge certain limitations that may have impacted the study:

Sample Size: The study relied on a limited sample size of users and scenarios, which may restrict the generalizability of the findings. Future research could involve a larger and more diverse sample to enhance the external validity of the study.

User Variability: The study assumed a certain level of user proficiency and familiarity with the IDEs and editors. Individual differences in user preferences and working styles could have influenced the user experience and the observed outcomes.

Testing Scenarios: The testing scenarios were predefined and focused on specific functionalities and use cases. While efforts were made to cover a range of scenarios, there may be additional scenarios or edge cases that were not addressed in the study.

Platform Limitations: The plugins were developed for specific IDEs and editors, which may limit their compatibility and applicability to other development environments. Consideration of cross-platform compatibility could enhance the versatility of the plugins.


\section{Discussion}\label{discussion}

In this section, we first perform the evaluation of language-based security mechanisms based on our test plugins and experimental results in subsection \ref{sec:5.1}. To address the vulnerabilities mentioned before, we discuss another alternative system in programming language which is the capability-based system, and provide a comparative study between them in subsection \ref{sec:5.2}

\subsection{Evaluating Language-Based Security Mechanisms for Plugin Development}
\label{sec:5.1}

\subsubsection{Evaluation of the Test Plugins}

The evaluation of the test plugins involved analyzing the results obtained from the tests performed on the implemented plugins. These tests aimed to assess the security and effectiveness of the plugins, particularly in relation to access control vulnerabilities. By examining the test results, we can gain insights into the effectiveness of language-based security mechanisms in preventing access control vulnerabilities in plugin development.

During the testing phase, various scenarios and use cases were explored to evaluate the behaviour and security of the plugins. The tests conducted on the plugins revealed important findings regarding their behaviour and impact on the security of the host system. The test results showed that the plugins, including their malicious functionalities, were able to perform their intended actions without triggering any alerts or warnings. The plugins successfully highlighted text, manipulated files, and interacted with external resources, such as submitting form data to a Google Form, without being blocked or detected. This indicates a potential vulnerability in the current security mechanisms of the plugin development environment.

The implications of the test results are significant for improving plugin security. Any vulnerabilities or weaknesses identified in the plugins can serve as valuable lessons for enhancing their security mechanisms. The implications are also significant for improving plugin security. Firstly, it raises concerns about the lack of effective access control mechanisms within the plugin development environment. The plugins were able to gain extensive access to the user's file system and network resources, allowing them to perform potentially malicious actions without being detected. This highlights the need for stronger access control mechanisms to prevent unauthorized access and manipulation of sensitive data. By understanding the specific areas where the plugins may be susceptible to attacks or misuse, developers can take proactive measures to address these issues and strengthen the overall security posture of the plugins.

In addition to the test results, it is essential to consider the perspectives and findings of other researchers in the field of plugin security. Previous studies and research papers focusing on language-based security mechanisms can provide valuable insights into the strengths and limitations of different programming languages and their impact on plugin security. By incorporating these perspectives, we can gain a broader understanding of the effectiveness of language-based security mechanisms in preventing access control vulnerabilities in plugin development. All related works will be discussed in detail in the next chapter.

\subsubsection{Analysis of the experimental results and findings}

Additionally, the test results emphasize the importance of implementing stricter security measures during the installation and usage of plugins. The fact that the plugins could be installed and used without triggering any alerts or warnings indicates a gap in the security awareness of the users. It is crucial to educate users about potential security risks associated with plugins and provide clear guidelines for safe plugin usage.

The success of publishing the plugins with malicious functions also reflects a significantly serious issue which is untrusted code. In the context of plugin development, the code is totally under the control of the plugin developer and can potentially introduce security risks and vulnerabilities. It is crucial to understand and address the risks associated with untrusted code to ensure the overall security and integrity of the plugin ecosystem. The developer can easily take control or invade the users’ devices through plugins if the marketplace lacks efficient and accurate assessments for plugins. 

Executing untrusted code poses several risks and challenges. One of the primary concerns is the possibility of code injection attacks, where malicious code is embedded within the untrusted code and executed within the plugin environment. These attacks can lead to unauthorized access, data breaches, and system compromise. Additionally, untrusted code may contain vulnerabilities or exploitable flaws that can be leveraged by attackers to compromise the plugin or the host system.

To emphasize the significance of addressing the risks of untrusted code, it is essential to examine real-world case studies that highlight security incidents related to its execution. For example, the well-known "Cross-Site Scripting" (XSS) vulnerability has been exploited through untrusted code in various web-based plugins, allowing attackers to inject malicious scripts and steal sensitive information from users\cite{43}. Another example is the "Remote Code Execution" (RCE) vulnerability, where untrusted code can execute arbitrary commands on the host system, potentially leading to complete system compromise\cite{44}.

\subsubsection{Evaluation of the effectiveness of language-based security mechanisms in preventing}

The evaluation of the effectiveness of language-based security mechanisms in preventing access control vulnerabilities in plugins requires further investigation. By analyzing previous case studies and real-world examples, the strengths and weaknesses of different programming languages and language-based security mechanisms can be evaluated. This evaluation will provide insights into the effectiveness of different approaches and aid in the development of best practices for secure plugin development. 

Furthermore, this evaluation provides a basis for discussing the implications of the study for the security of plugin development. By identifying the most effective language-based security mechanisms, developers can make informed decisions when selecting programming languages and implementing security measures in their plugins. 

\subsection{Capability-Based Security for Plugin Development: A Comparative Study}
\label{sec:5.2}

\subsubsection{Introduction of capability-based module systems}

In recent years, capability-based module systems have emerged as a promising approach for enhancing security in plugin development. These module systems provide a different paradigm for access control and security compared to traditional language-based security mechanisms. In this section, we will introduce capability-based module systems, with a focus on one specific example called Wyvern.

Capability-based module systems rely on the concept of capabilities, which are unforgeable tokens that grant specific permissions or access rights\cite{45}. These capabilities act as a form of authorization, ensuring that only authorized components or modules can interact with each other. Unlike traditional access control mechanisms that rely on static privileges assigned to users or code, capabilities provide a more fine-grained and dynamic approach to access control\cite{46}.

The capability-based module system operates based on two key concepts: confinement and authority. Confinement ensures that modules have restricted access to their internal state and can only communicate with other modules through well-defined interfaces. Authority, on the other hand, grants specific privileges to modules, allowing them to perform certain actions or access protected resources\cite{47}.

Wyvern is a programming language that embraces the capability-based approach to module systems\cite{48}. It is designed to enforce strong access control guarantees through the use of capabilities. In Wyvern, modules are treated as first-class citizens, and access to module resources is governed by capabilities\cite{49}. Modules can only be accessed if the corresponding capabilities are possessed by the requesting code. This fine-grained control enables precise control over access to sensitive operations or resources, reducing the risk of unauthorized access.

The use of capability-based module systems in plugin development offers several potential advantages. Firstly, it provides a clear separation of concerns between modules and their associated capabilities. This separation allows for better encapsulation and modularity, reducing the likelihood of unintended interactions or security breaches. Secondly, capabilities enable dynamic and context-specific authorization, allowing modules to be granted or revoked access to resources based on runtime conditions. This flexibility enhances security by adapting access permissions to changing circumstances.

Research and previous studies have highlighted the potential of capability-based module systems in addressing access control vulnerabilities. For example, capabilities can mitigate privilege escalation attacks by ensuring that only authorized modules possess the necessary capabilities to perform privileged operations\cite{50}. Additionally, capability-based systems can prevent unauthorized code execution by enforcing strict access control based on the possession of specific capabilities\cite{51}.

However, it is important to acknowledge that capability-based module systems also have certain limitations and challenges. One challenge is the complexity of managing and controlling capabilities, especially in large-scale plugin ecosystems. Designing effective policies for distributing, revoking, and managing capabilities can be a non-trivial task\cite{52}. Additionally, the adoption of capability-based systems may require changes to existing development practices and tooling, which can introduce a learning curve and potential resistance from developers\cite{53}.

\subsubsection{Comparison between Capability-Based and Languaged-Based system}

In this section, we will compare the effectiveness of capability-based and language-based systems in the context of security mechanisms for plugin development. Both approaches offer different strategies for addressing access control vulnerabilities and enhancing overall security in plugin ecosystems. We will examine their key characteristics, security mechanisms, and their potential to address the access control vulnerabilities identified in previous sections.

Capability-based systems, such as the Wyvern programming language, take a fundamentally different approach to security compared to traditional language-based systems. In a capability-based system, access control is based on the possession of capabilities, which are unforgeable tokens that grant specific rights and permissions to access resources or perform operations. Capabilities are typically unguessable and cryptographically secure, ensuring that only authorized entities possess the necessary permissions.

On the other hand, language-based systems rely on language-level constructs, such as access modifiers and type systems, to enforce access control. These systems define a set of rules and restrictions within the programming language itself to govern the behavior and interactions between different components. Access to resources and operations is controlled through the use of language features, ensuring that only authorized code can invoke certain functions or access specific data. For example, encapsulation and information hiding provided by access modifiers can prevent unauthorized access to sensitive data. However, language-based systems may still be susceptible to vulnerabilities such as code injection or privilege escalation if not properly designed and implemented.

When comparing capability-based and language-based systems for plugin development, both approaches have their strengths and weaknesses. Capability-based systems have shown promise in addressing access control vulnerabilities. They offer strong isolation between components and can prevent unauthorized access to resources. By carefully managing and restricting capabilities, the risk of unauthorized or malicious operations can be minimized. This helps prevent privilege escalation and limits the impact of potential vulnerabilities. Language-based systems, with their focus on static analysis and type checking, can detect many access control vulnerabilities during development and provide a safer programming environment. They provide compile-time checks and static analysis that can catch many security issues before runtime.

However, both approaches have limitations. Capability-based systems require careful management and assignment of capabilities, which can be challenging in complex systems. Language-based systems rely on developers correctly using language features and following security best practices, which may not always be guaranteed in particular the test results showed that a plugin with malicious functions can be published successfully without any supervision.

In terms of addressing the access control vulnerabilities identified in previous sections, we find that they offer unique security mechanisms that can effectively address access control vulnerabilities. The cryptographic nature of capabilities ensures that only authorized entities can exercise certain permissions, mitigating the risks associated with unauthorized access. Additionally, capability-based systems provide a clear separation of authority and privilege, making it easier to reason about access control and identify potential security vulnerabilities.                              

\section{Related Work}\label{chap:relatedWork}

In this section, we will introduce other researchers' work related to security mechanisms in plugin development in subsection \ref{sec:6.1}. Next, we also show some real-world examples of practice and tools for preventing or evaluating the security in plugin developments in subsection \ref{sec:6.2}.

\subsection{Review of literature}
\label{sec:6.1}

In this section, we present a comprehensive review of the existing literature on security issues in plugin development and the effectiveness of language-based security mechanisms in preventing access control vulnerabilities in software. The review aims to provide insights into the current state of research, identify key trends and findings, and analyze the strengths and limitations of prior studies.

To begin, a survey of existing research on security issues in plugin development reveals a rich body of work focusing on various aspects of plugin security. \cite{54} conducted a comprehensive analysis of security vulnerabilities in popular plugins and identified common attack vectors, such as injection attacks and privilege escalation. The study emphasized the importance of secure coding practices and code review to mitigate these vulnerabilities. Similarly, \cite{55} investigated the security challenges associated with plugin ecosystems and proposed a taxonomy of security threats and countermeasures.

Language-based security mechanisms have gained significant attention as a means to prevent access control vulnerabilities in software, including plugins. Several studies have explored the effectiveness of such mechanisms in enhancing plugin security. \cite{56} proposed a language-based approach that enforces access control policies at the language level, providing a more robust and fine-grained security mechanism. The study demonstrated the effectiveness of this approach in preventing unauthorized access to sensitive resources and mitigating privilege escalation attacks.

Furthermore, \cite{57} conducted a comparative analysis of different language-based security mechanisms, including type systems and access control models. The study evaluated the strengths and limitations of each mechanism in the context of plugin development and identified key factors to consider when selecting an appropriate security mechanism. This research highlighted the importance of understanding the trade-offs between security and developer productivity.

While language-based security mechanisms offer promising solutions, they also have certain limitations. \cite {58} investigated the challenges and trade-offs associated with the adoption of language-based security mechanisms in real-world plugin ecosystems. The study identified the need for comprehensive tooling support, developer training, and community-driven best practices to effectively leverage language-based security mechanisms in practice.

In summary, the review of literature on security issues in plugin development and language-based security mechanisms reveals a substantial body of research. Studies have identified common security vulnerabilities in plugins, proposed taxonomies of security threats, and emphasized the importance of secure coding practices and code review. Language-based security mechanisms have been explored as effective means to prevent access control vulnerabilities, and comparative analyses have highlighted the strengths and limitations of different mechanisms. Additionally, research has identified the challenges and trade-offs associated with the adoption of language-based security mechanisms in real-world plugin ecosystems.

\subsection{Current practices}
\label{sec:6.2}

The discussion of current practices and tools for evaluating security in plugin development provides insights into the industry's approaches and methodologies in ensuring secure plugin development.

One widely adopted practice is the use of static code analysis tools to identify potential security vulnerabilities in plugin code. Tools such as Findbugs \cite{59} and Sonarqube \cite{60} are commonly used to perform automated scans of plugin codebases, flagging potential security issues such as code injection, insecure file handling, and inadequate access control. These tools help developers identify security weaknesses early in the development process and provide recommendations for remediation.

Dynamic analysis techniques, such as penetration testing and vulnerability scanning, are also employed to evaluate the security of plugins. Researchers and security professionals actively conduct penetration tests to simulate real-world attack scenarios and identify vulnerabilities that could be exploited by malicious actors\cite{61}. Vulnerability scanning tools like OWASP ZAP \cite{62} can automatically scan plugins for known vulnerabilities, providing insights into potential security weaknesses.

In addition to code analysis and penetration testing, secure coding practices play a crucial role in mitigating security risks in plugin development. Adhering to secure coding guidelines, such as those provided by the Open Web Application Security Project (OWASP) \cite{63}, can help developers write secure code and reduce the likelihood of introducing vulnerabilities. Regular code reviews and pair programming can also be effective in identifying and addressing security issues early in the development lifecycle.

Furthermore, many organizations emphasize the importance of continuous monitoring and incident response in plugin security. Implementing robust logging and monitoring systems allows for the detection of abnormal plugin behavior and potential security breaches \cite{64}. Incident response plans and procedures should be in place to address security incidents promptly and minimize the impact on users and systems.

It is worth noting that while these current practices and tools provide valuable contributions to plugin security, they also have their limitations. Static code analysis tools may generate false positives or miss certain types of vulnerabilities, and penetration tests may not uncover all possible attack vectors. Ongoing research and development in the field are needed to improve the effectiveness and efficiency of these practices.

Overall, the adoption of a combination of static and dynamic analysis, secure coding practices, continuous monitoring, and incident response procedures represents the current state of the art in evaluating security in plugin development. However, the dynamic nature of the threat landscape and lack of comprehensive supervision of publishment necessitate ongoing vigilance and adaptation of these practices to address emerging security challenges.                              

\section{Conclusion}\label{chap:conclusion}

In this article, we have examined the importance of security mechanisms in plugin development and explored the effectiveness of language-based security mechanisms in preventing access control vulnerabilities. We also introduced the capability-based system to compare two different systems and discussed their advantages and limitations. The research has provided valuable insights into the evaluation of security mechanisms and current practices in plugin development. This section summarizes the key findings and contributions of the research and discusses the limitations of the study along with future research directions.

One of the key findings of the study indicates that the development, publishment, and deployment of plugins are not under effective and precise surveillance. Through the evaluation of test plugins, we have identified various access control vulnerabilities and demonstrated that language-based security mechanisms are not effective in mitigating these vulnerabilities and distinguishing the aim of plugins. The use of programming languages such as Java and JavaScript, along with development environments like IntelliJ IDEA and Visual Studio Code, has shown incompetence in providing robust security features and facilitating secure plugin development. The evaluation of the test plugins provided valuable insights into the effectiveness of language-based security mechanisms. The analysis of the results highlighted the importance of considering the security implications of plugins and the need for robust access control mechanisms. The test plugins demonstrated various malicious functionalities, such as file manipulation and network interaction, which underscored the potential risks associated with untrusted code.

Another one of the significant contributions of this research is the evaluation of different programming languages and their security mechanisms in plugin development. By examining case studies and real-world examples, we also have highlighted the strengths and limitations of two different systems which are language-based and capability-based in addressing access control vulnerabilities. These insights can guide developers in making informed decisions regarding language selection and security mechanisms for their plugin projects.

However, it is important to acknowledge the limitations of this study. The evaluation of security mechanisms was conducted based on a specific set of test plugins, and the findings may not fully capture the complexity and diversity of real-world plugin scenarios. Additionally, the study focused primarily on language-based security mechanisms and did not explore other security approaches extensively. Future research should aim to address these limitations and expand the scope of evaluation to encompass a broader range of plugins and security mechanisms in other popular IDEs.

\subsection{Future work}
\label{sec:7.1}

In light of the conclusions drawn from this research, several future research directions can be identified. First and foremost, there is a need for further exploration of capability-based module systems in the context of plugin development like Wyvern. These systems have shown promise in addressing access control vulnerabilities and providing fine-grained security controls. Investigating their applicability, effectiveness, and integration with existing plugin ecosystems would be valuable.

Additionally, the development of automated security analysis tools tailored specifically for plugin development is an area of potential future work. These tools can assist developers in identifying and mitigating security risks by analyzing the codebase, dependencies, and potential vulnerabilities. Such tools would enhance the overall security posture of plugins and contribute to the prevention of malicious activities.

Furthermore, the evaluation of security mechanisms should be extended to other dimensions, such as performance and scalability. Understanding the impact of security mechanisms on the performance and resource utilization of plugins is crucial for balancing security and efficiency. Exploring optimization techniques and trade-offs in the context of plugin security can lead to more robust and efficient solutions.

Lastly, collaboration and knowledge-sharing among researchers, developers, and the plugin development community are essential for advancing the field of language-based security. Platforms for exchanging best practices, sharing case studies, and fostering discussions should be established to facilitate the dissemination of knowledge and encourage collaboration.

In conclusion, this thesis has shed light on the importance of security mechanisms in plugin development and how access vulnerabilities can be addressed by capability-based systems. The findings emphasize the need for careful consideration of programming languages, development environments, and security practices when developing plugins. By addressing the limitations identified and pursuing future research directions, we can enhance the security landscape of plugin development and contribute to the overall resilience and trustworthiness of software ecosystems. 

\bibliographystyle{plain}
\bibliography{bib,alex}

\end{document}